\documentclass{INTERSPEECH2024}

\usepackage{caption}
\usepackage{subcaption}
\usepackage{url}
\usepackage{hyperref}
\usepackage{algorithm}
\usepackage{algpseudocode}
\usepackage{graphicx}
\usepackage{titlepic}
\usepackage{multicol, multirow}

\makeatletter
\def\BState{\State\hskip-\ALG@thistlm}
\makeatother

\interspeechcameraready

\title{\textbf{Audio Enhancement from Multiple Crowdsourced Recordings: \\ A Simple and Effective Baseline}}

\author{\emph{Shiran Aziz ~~~~~~~~~~ Yossi Adi ~~~~~~~~~~ Shmuel Peleg} \\ 
School of Computer Science and Engineering\\ \vspace{0.1cm}
The Hebrew University of Jerusalem, Israel\\
\small\url{https://shiranaziz.github.io/crowdsourced_audio_enhancement/} \\ \vspace{0.1cm}
\small\texttt{shiran.aziz@mail.huji.ac.il}
}
\date{}


\titlepic{\includegraphics[width=\textwidth]{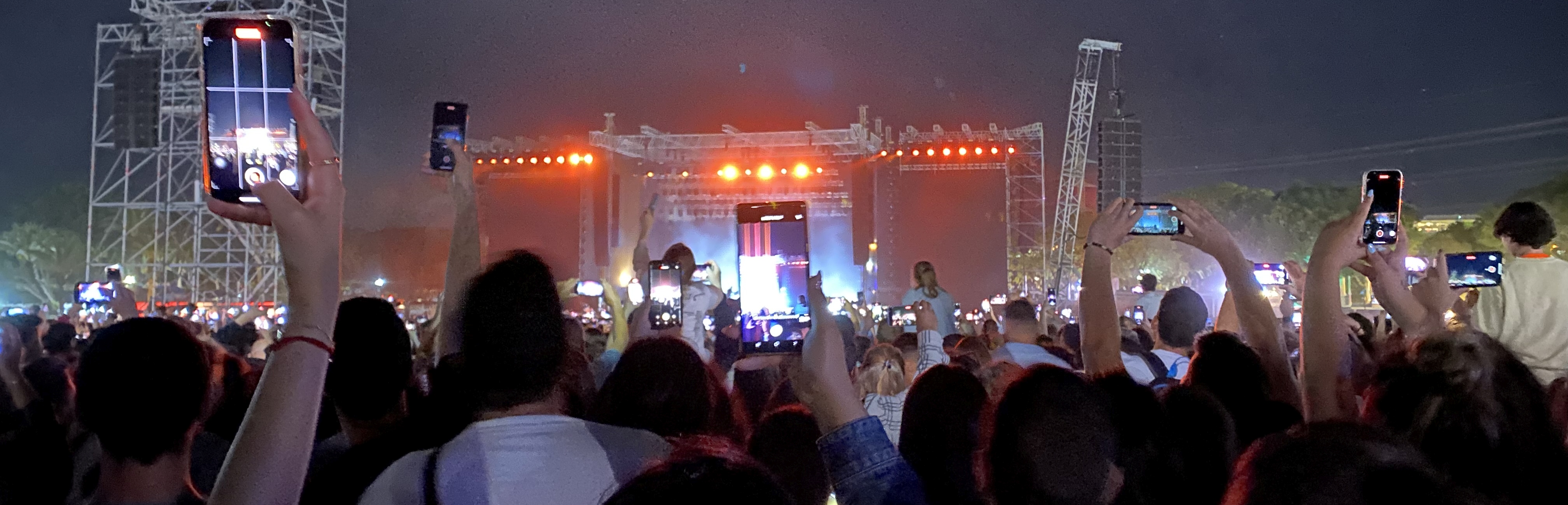}
\captionof{figure}{A rock concert recorded with multiple cameras by the audience.\label{fig:concert}}}

\begin{document}

\maketitle    

\begin{abstract}
With the popularity of cellular phones, events are often recorded by multiple devices from different locations and shared on social media. Several different recordings could be found for many events. Such recordings are usually noisy, where noise for each device is local and unrelated to others. This case of multiple microphones at unknown locations, capturing local, uncorrelated noise, was rarely treated in the literature. In this work we propose a simple and effective crowdsourced audio enhancement method to remove local noises at each input audio signal. Then, averaging all cleaned source signals gives an improved audio of the event. We demonstrate the effectiveness of our method using synthetic audio signals, together with real-world recordings. 
This simple approach can set a new baseline for crowdsourced audio enhancement for more sophisticated methods which we hope will be developed by the research community.   
\end{abstract}

\section{Introduction}
\label{sec:intro}

Cellular phones are powerful multimedia devices, capable of quality recording of events around us. In particular, public events are often captured by multiple people from different locations. See Fig~\ref{fig:concert} for a sample rock concert. Many such user-generated recordings are also uploaded to social media, where several different recordings could be found for each event. In most cases user recordings have noisy audio signals, where noises are mostly local to each device, and unrelated to each other due to the distance between users. Crowdsourced audio enhancement aims to use all available audio signals of an event, creating an audio signal that excludes the local noises at each input signal.

Unlike more traditional single-channel and multi-channel denoising approaches~\cite{ephraim1984speech, nugraha2016multichannel, gannot2017consolidated, defossez2020real, araki2019projection, chazan2021single}, in crowdsourced audio enhancement there is no prior definition of noise. Instead, noise is defined as a sound that is not common to most input audio. Hence, while local sounds will be removed, any global sounds that are present in all input signals will remain. For instance, consider several people shooting with their cell phones videos of a musical concert from different locations in the hall. The music coming from the main stage will be captured in all recordings, however the background noise will be unique to each of the recordings. 

This work presents a straightforward method for crowdsourced audio enhancement. The method is based on filtering noisy space-time outliers from the input spectrograms considering both upper and lower thresholds. Specifically, we start by computing the Short-Time Fourier Transform (STFT) of all input signals. 
For each Time-Frequency (TF) cell we examine the magnitude values given to it by each input signal, and outlier values in each cell are removed. We define outliers as values which are substantially higher or lower than the median magnitude of the corresponding TF cell. The enhanced signal is constructed by averaging all STFT in each TF cell that are not outliers. We evaluated the proposed method considering both synthetic and in-the-wild recordings. Results suggest that the proposed method significantly outperforms the baseline methods considering a diverse set of sources and background noises. 
The proposed method is simple and straightforward, requires no training, hence can serve as a foundational baseline for comparison with more sophisticated statistical techniques.

\section{Related Work}

\label{sec:related_works}
While much work has been done on audio enhancement using multi-channel microphone arrays~\cite{benesty2017fundamentals}, most papers are \textit{position-aware} and address the case where the properties of the microphones and the relationship between them are known and constant~\cite{vincent2018audio,tzirakis2021multi,pandey2022multichannel}.

Combining user recordings should be \textit{position-agnostic}, as we do 
not have any prior information on the relative position of the microphones. The authors in~\cite{Kim2013collaborative} were the first, to the best of our knowledge, to address crowdsourced audio enhancement from unrelated recordings. They proposed creating an improved audio signal, where the possible corruptions in each input signal can be missing frequencies or missing time periods. 

Another relevant line of work is \textit{scene-agnostic} multi-microphone speech processing.
The authors in~\cite{Yemini2010derever} proposed a deep learning based solution for speech dereverberation considering a varying number of microphone array at different positions. 
Some papers \cite{yoshioka2022picknet,taherian2022one} are focused on a setup where the target speaker is always closest to the microphone array. Unlike this approach, we have a single clean source, and we can not assume that one microphone has the cleanest recording of this source. Recently, \cite{jukic2023flexible} showed, in parallel to our work, a flexible multichannel speech enhancement, for a varying number of microphones at random positions inside a room. Though they show impressive results they focus on indoor speech recordings with relatively small distances between the microphones in the array.
Unlike the crowdsourced speech enhancement task, these lines of work assume that all sources are 
captured by all microphones. Similarly, in Independent Component Analysis (ICA)~\cite{hyvarinen2000independent} multi-channel speech separation is done by finding a linear representation of non-Gaussian data so that the components are as statistically independent as possible. Notice, ICA considers equal number of sources and microphones. Following such line of research the authors in~\cite{duong2010under} proposed the \emph{Full-rank spatial Covariance
Analysis} (FCA) method, while the authors in~\cite{ito2021joint} proposed the \emph{fastFCA}, which extends such research direction and proposed a method for source separation for the undetermined case of more sources than microphones. 

In this work, we address the case where each audio signal has an independently added noise. Similarly to our setup, the authors in~\cite{MaxElimination:2017} propose the \emph{Max-elimination} method, which removes at each time-frequency cell the signal having a maximal amplitude. This is the most similar approach to our method, and when comparing our results to this method, and find that our results are better.

\section{Crowdsourced Audio Enhancement}

We address an audio source $S$, recorded by $m$ independent microphones at unknown locations. Let $A_1, ..., A_m$ be the input signals from each of the microphones, where each signal $A_i$ is composed of the source signal $S$ at some time period, together with added noise $N_i$.  It is assumed that microphones are far from each other such that all $m$ noises are different and uncorrelated. Our proposed method starts by temporally aligning all input signals and normalizing their magnitude. Then, we denoise the input signals using time-frequency filtering. 

\begin{figure}[t!]
  \centering
  \includegraphics[width=\columnwidth]{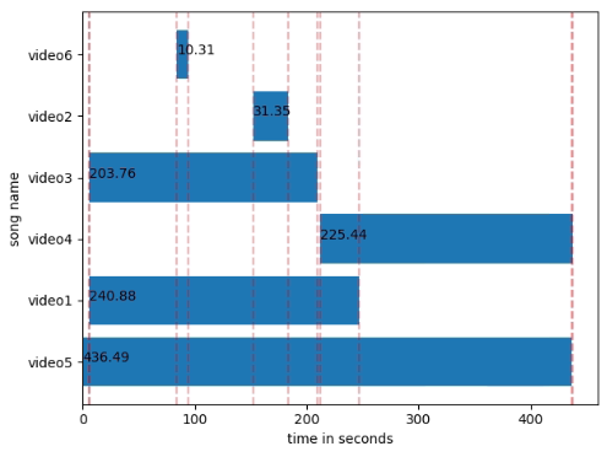}
  \caption{Clips after temporal alignment. For each time period there may be different clips covering this period. In this example we have 5 input clips, where some periods are covered by 1, 2, 3, or 4 simultaneous clips.}
  \label{fig:timeline}
\end{figure}

For temporal alignment of the audio signals we use the method proposed by~\cite{wang2003industrial}, using time-frequency magnitude peaks. Alignment is done by finding correspondences between frequencies and time differences of detected pairs of peaks\footnote{we use the implementation from \url{https://github.com/worldveil/dejavu}}. Aligned clips are shown on a timeline in Fig.~\ref{fig:timeline}, where we see that for each time period we may have a different number of overlapping clips. After temporal alignment, the corresponding peaks used in the alignment process are assumed to belong to the clean audio source, and the amplitude at the corresponding peaks are normalized accordingly. We normalize by estimating the multiplicative constant between all corresponding pairs. We first select the signal with the maximum number of matched peaks as an anchor signal, and normalize the other signals by multiplying them by a corresponding $\alpha$ value for each signal.
Formally, given are pairs of matched spectral peaks in the log-spectrogram,
$\{(|P^X_n|, |P^Y_n|)\}_{n=1}^{N}$
where $|P_n^X|$ is the amplitude of the $n^{th}$ frequency peak of signal $X$ and $|P_n^Y|$ is the corresponding peak in $Y$. Using the log spectrogram amplitude, we estimate the coefficient from the mean of all the pairs as 
$\alpha^{XY} = (\sum_{n=1}^N |P_n^X|) / \sum_{n=1}^N (|P_n^Y|)$.

\begin{figure}[tbh!]
  \centering
  \includegraphics[width=\columnwidth, height=0.5\columnwidth]{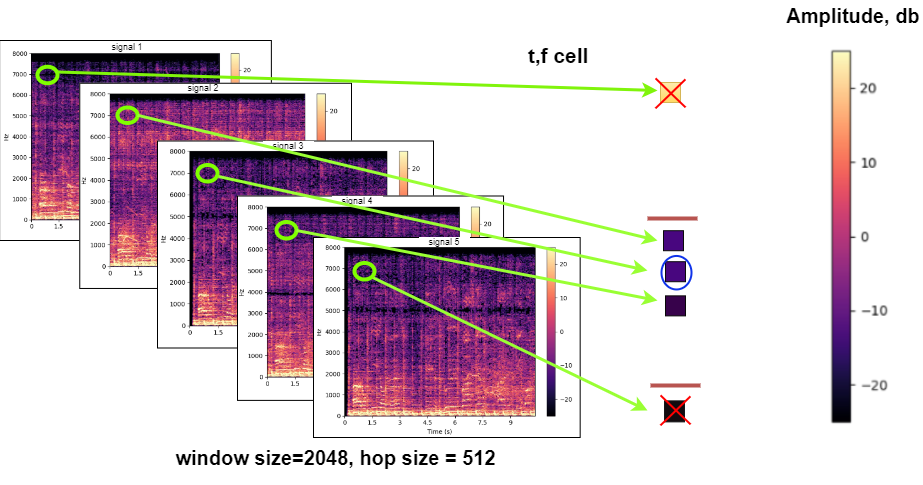}
  \caption{The audio enhancement process: For each TF cell in the spectrogram of overlapping clips we examine the amplitudes in each clip, compute the median amplitude, and remove values whose distance from the median exceeds a threshold. Averaging the complex values from the remaining clips give the value of TF cell in the enhanced spectrogram. In this figure we have 5 overlapping clips, and of the 5 amplitudes in the examined TF cell the highest and lowest amplitudes are discarded as outliers.}
  \label{fig:filter}
\end{figure}

Once all signals $A_1, ..., A_m$ are aligned and normalized, 
we estimate the source signal $S$ at each time $t$ from all input signals available at $t$. This is done by removing outliers at each $(t,f)$ cell. Formally, for all input signals $A_i$, we compute the complex STFTs $Y_i$, using $2048$ FFT coefficients, window size of $2048$, and overlap ratio of $0.25$. Next, for each $(t,f)$ cell we perform the following: (i) given all magnitude STFT $|Y_i|$ defined at time $t$ compute the median amplitude in the $(t,f)$ cell, denoted as $C(t,f)$; (ii) define as outliers those signals whose STFT magnitudes are above or below given thresholds, that depend on the median $C(t,f)$. Formally, outlier values are those that satisfy 
$|Y_i(t,f)| > \lambda_1 C(t,f)$ or 
$|Y_i(t,f)| < \lambda_2 C(t,f)$, where $\lambda_1$ and $\lambda_2$ are hyper-parameters calibrated on the available dataset; (iii) The denoised complex STFT, $G(t,f)$ is constructed by 
averaging the values of all signals that were not detected as outliers. 
Intuitively, when $|Y_i(t,f)|$ is substantially larger than the median or substantially lower than the median of all input signals, it is considered as noise. Next, we relax the the prior outlier criteria: We examine TF cells in the neighborhood of a removed cell, and also remove those values that fulfill a relaxed outlier threshold $\gamma$ (instead of $\lambda_1$). 
If all signals in a cell are removed by the above process, only the upper threshold is used. Lastly, we convert $G(t,f)$ back to a time-domain signal by applying inverse STFT using the mean phase of all signals. In all experiments we use $\lambda_1=1.15$, $\gamma = 1.1$ and $\lambda_2 = 0.01$. A pseudo-code of the proposed algorithm can be found at Algorithm~\ref{alg:method}.

\begin{algorithm}[tbh!]
\caption{Filtering a time segment having $k$ overlapping signals
\label{alg:method}}
\begin{algorithmic}[1]
\For {$i=1,2,\ldots k$}
\State $Y_i = STFT(A_i)$
\State $M_i \leftarrow \text{ones of the shape } Y_i$
\EndFor
\State $C \leftarrow median(\{|Y_1|,...,|Y_k|\})$
\For {$i=1,2,\ldots k$}
\For{$(t,f)$ in $Y_i$}:
\If{ $|Y_i(t,f)|>\lambda_1 C(f,t)$ or \par
~~~\hskip\algorithmicindent $|Y_i(t,f)|< \lambda_2 C(f,t)$} \par ~~~~~~~~~\hskip\algorithmicindent $M_i(f,t)\leftarrow 0$
\EndIf
\EndFor
\State $G_i^0 = Y_i$, \ $ G_i^1 \leftarrow M_i \odot Y_i$
\While {$G_{i}^{j-1} \neq G_i^{j}$} 
\For {$(t,f)$  which $M_i(t, f) = 0$}
\For{ $t-1 \leq s \leq t+1$ and \par
\hskip\algorithmicindent ~~~~~~~~~~~~~~~~~~ $f-1 \leq g \leq f+1$}
    \If{$|Y_i(s,g)| > \gamma C(s,g)$} \par ~~~~~~~~~~~~~~~~~~\hskip\algorithmicindent $M_i(s, g) \leftarrow  0$ 
    \EndIf
    \State update $G_i^j \leftarrow M_i \odot Y_i$
   \EndFor
    \EndFor
\EndWhile
\EndFor
\State 
$G \leftarrow (\sum_{i=1}^{k} G_i^{final}) ~/~ (\sum_{i=1}^{k}M_i) $
\State \Return $ISTFT(G)$
\end{algorithmic}
\end{algorithm}
\begin{figure*}[t!]
     \centering
     \begin{subfigure}[b]{0.32\textwidth}
         \centering
         \includegraphics[width=0.9\columnwidth, height=0.7\columnwidth]{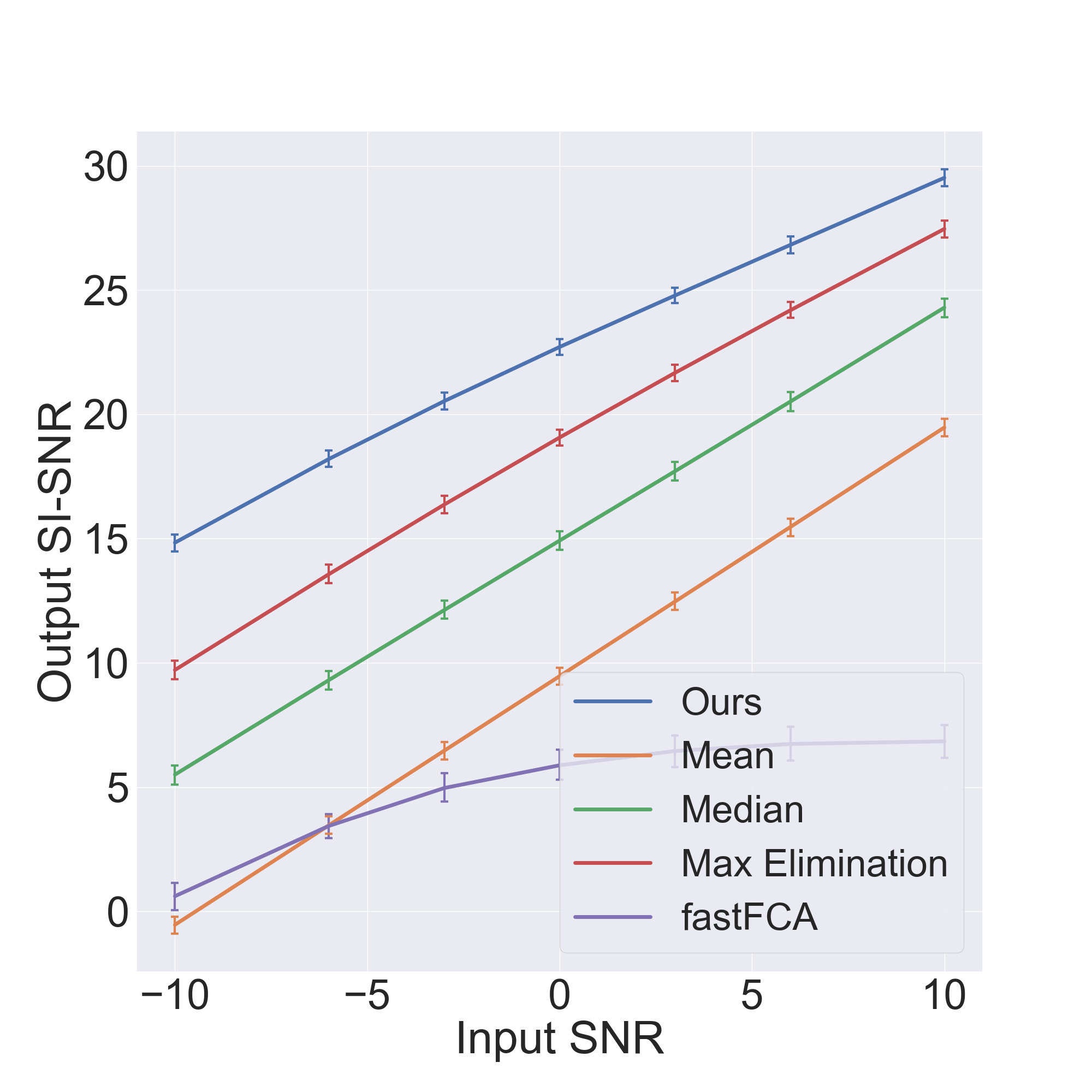}
         \caption{Singal: Music vs. Noise: Speech\label{fig:synthetic_music_spk}}
     \end{subfigure}~~~~
     \begin{subfigure}[b]{0.32\textwidth}
         \centering
         \includegraphics[width=0.9\columnwidth, height=0.7\columnwidth]{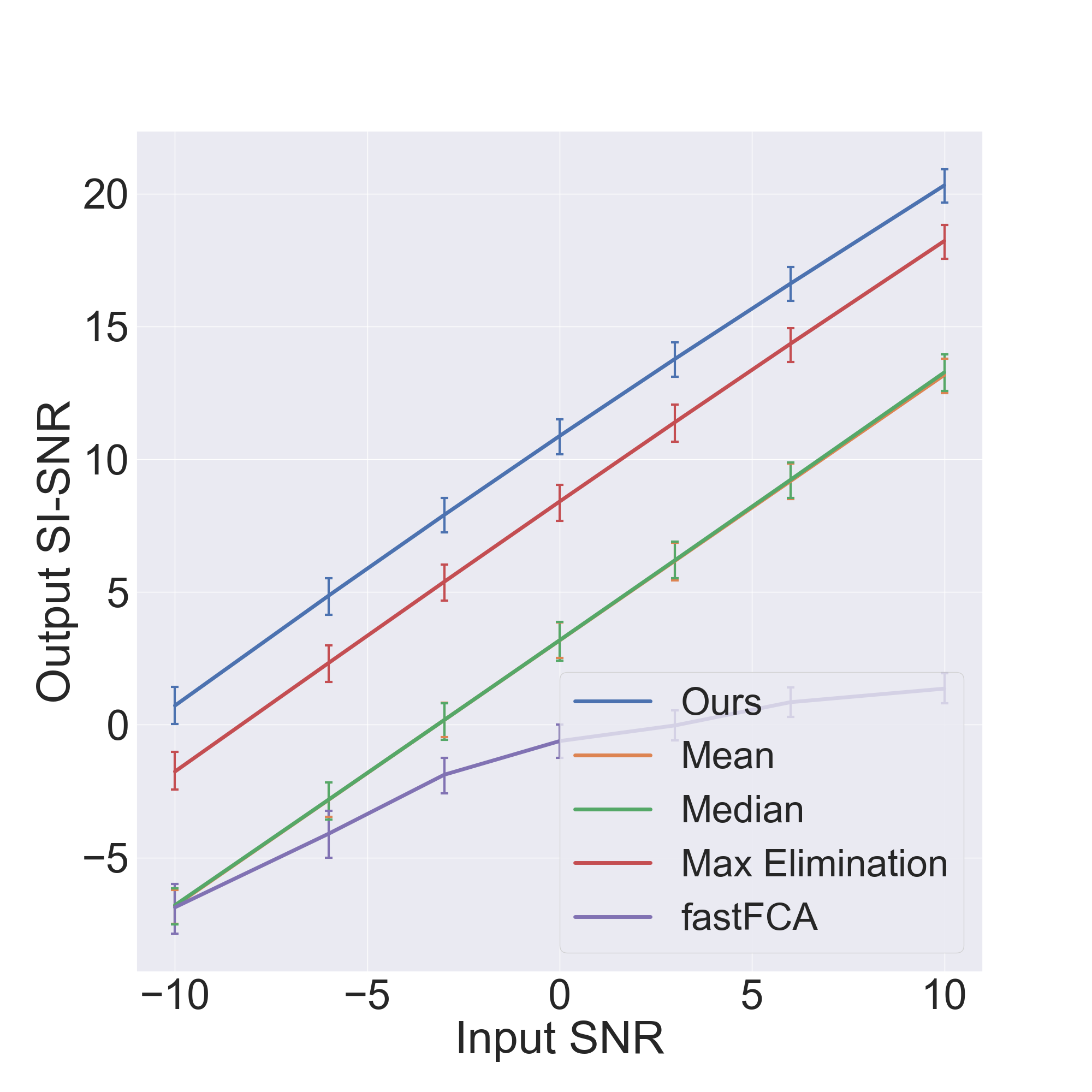}
         \caption{Music vs. Noises (DEMAND)\label{fig:synthetic_music_demand}}
     \end{subfigure}~~~~
     \begin{subfigure}[b]{0.32\textwidth}
         \centering
         \includegraphics[width=0.9\columnwidth, height=0.7\columnwidth]{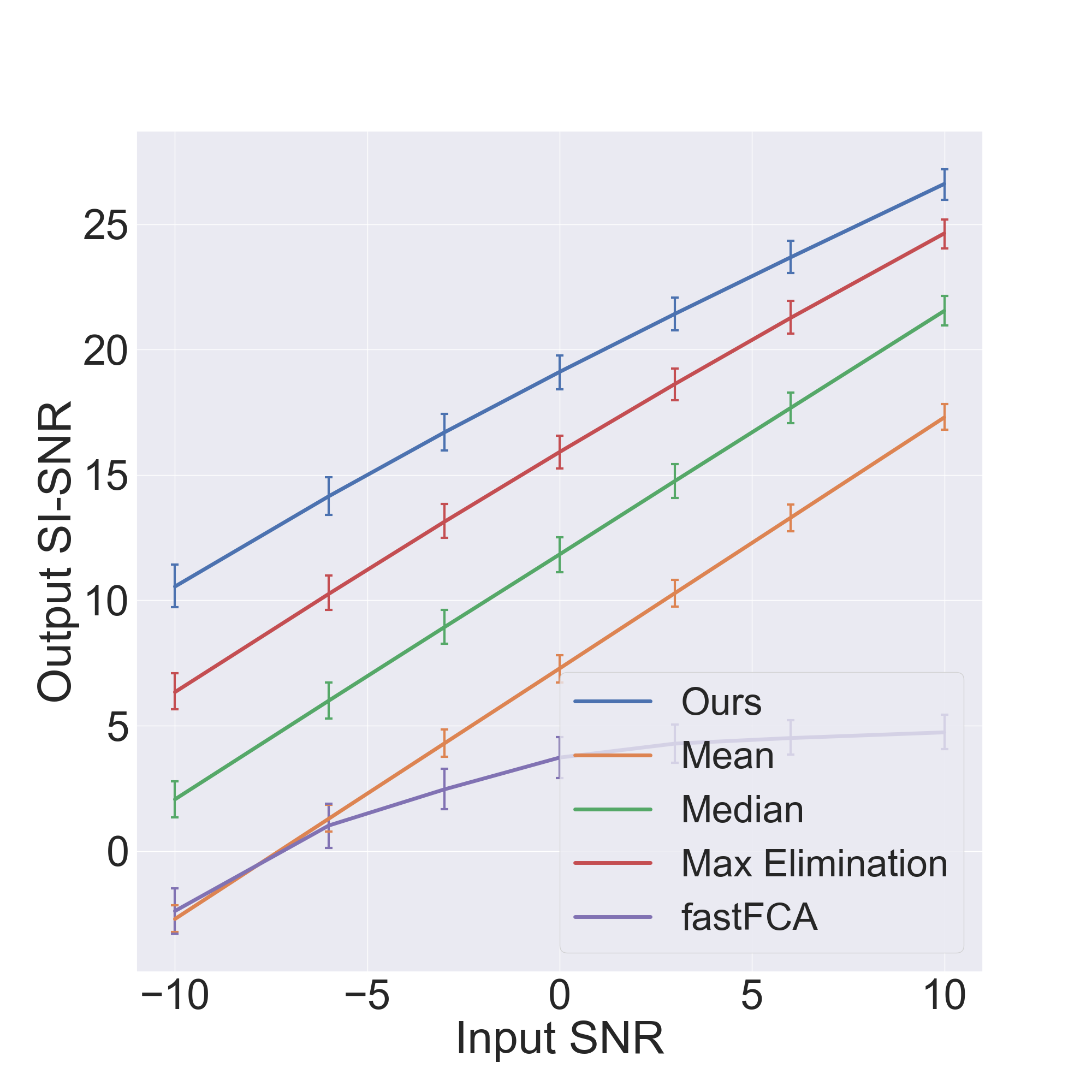}
         \caption{Music vs. Noises (DNS) \label{fig:synthetic_music_dns}}
     \end{subfigure}
     \\
       \begin{subfigure}[b]{0.32\textwidth}
         \centering
         \includegraphics[width=0.9\columnwidth, height=0.7\columnwidth]{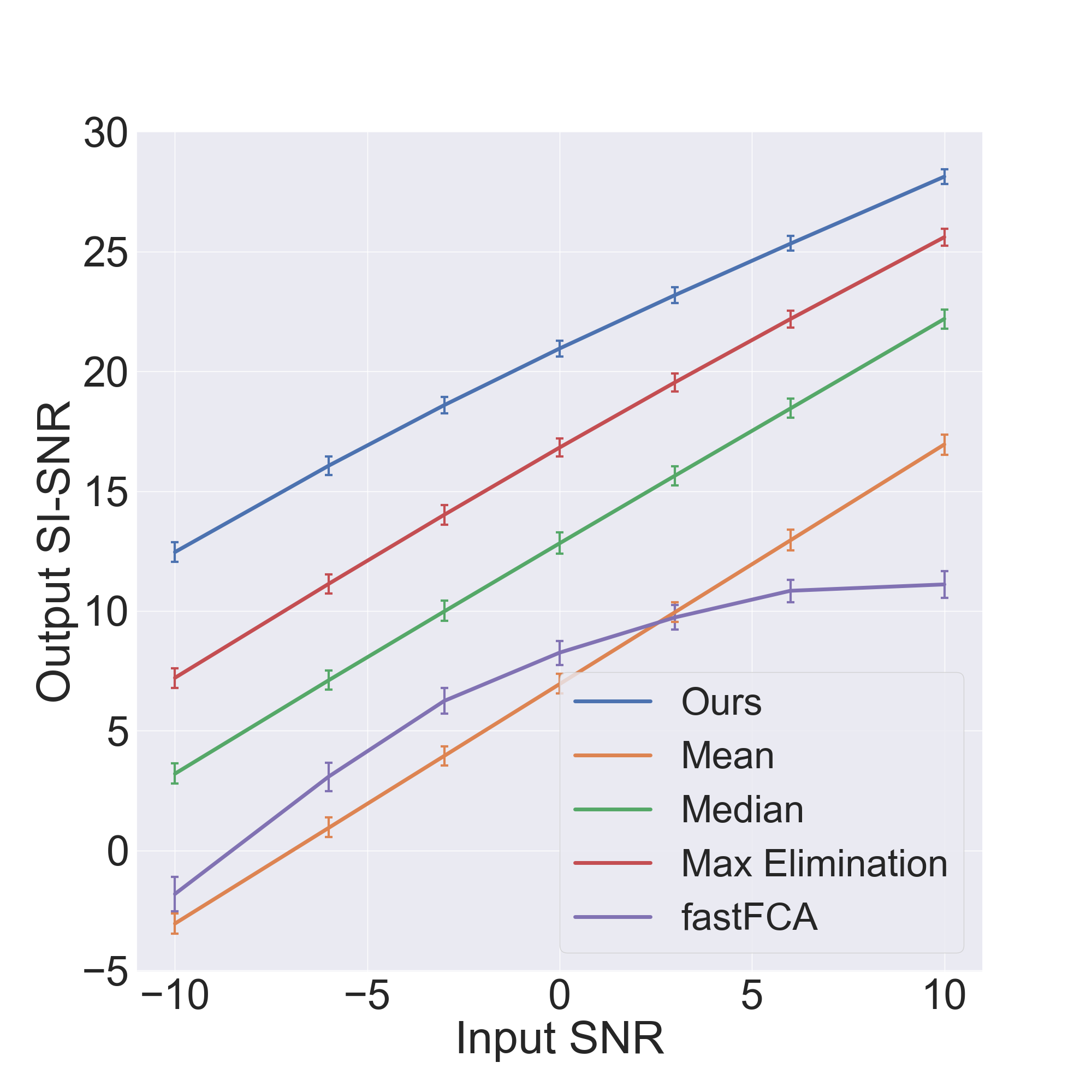}
         \caption{Signal: Speech vs. Noise: Speech \label{fig:speech_and_different_speech_noises}}
     \end{subfigure}~~~~
     \begin{subfigure}[b]{0.32\textwidth}
         \centering
         \includegraphics[width=0.9\columnwidth, height=0.7\columnwidth]{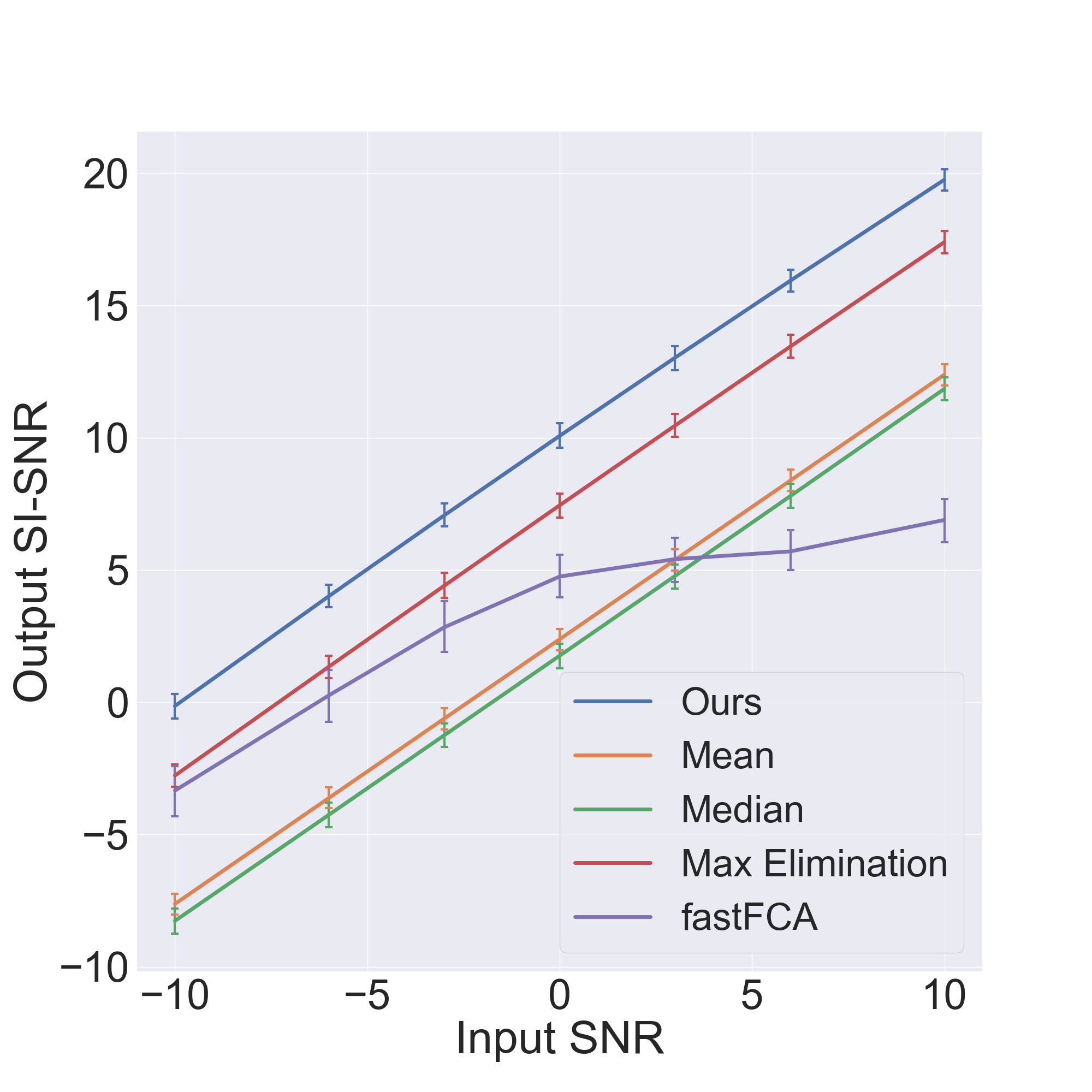}
         \caption{Speech vs. Noises (DEMAND) \label{fig:speech and different environmental noises}}
     \end{subfigure}~~~~
     \begin{subfigure}[b]{0.32\textwidth}
         \centering
         \includegraphics[width=0.9\columnwidth, height=0.7\columnwidth]{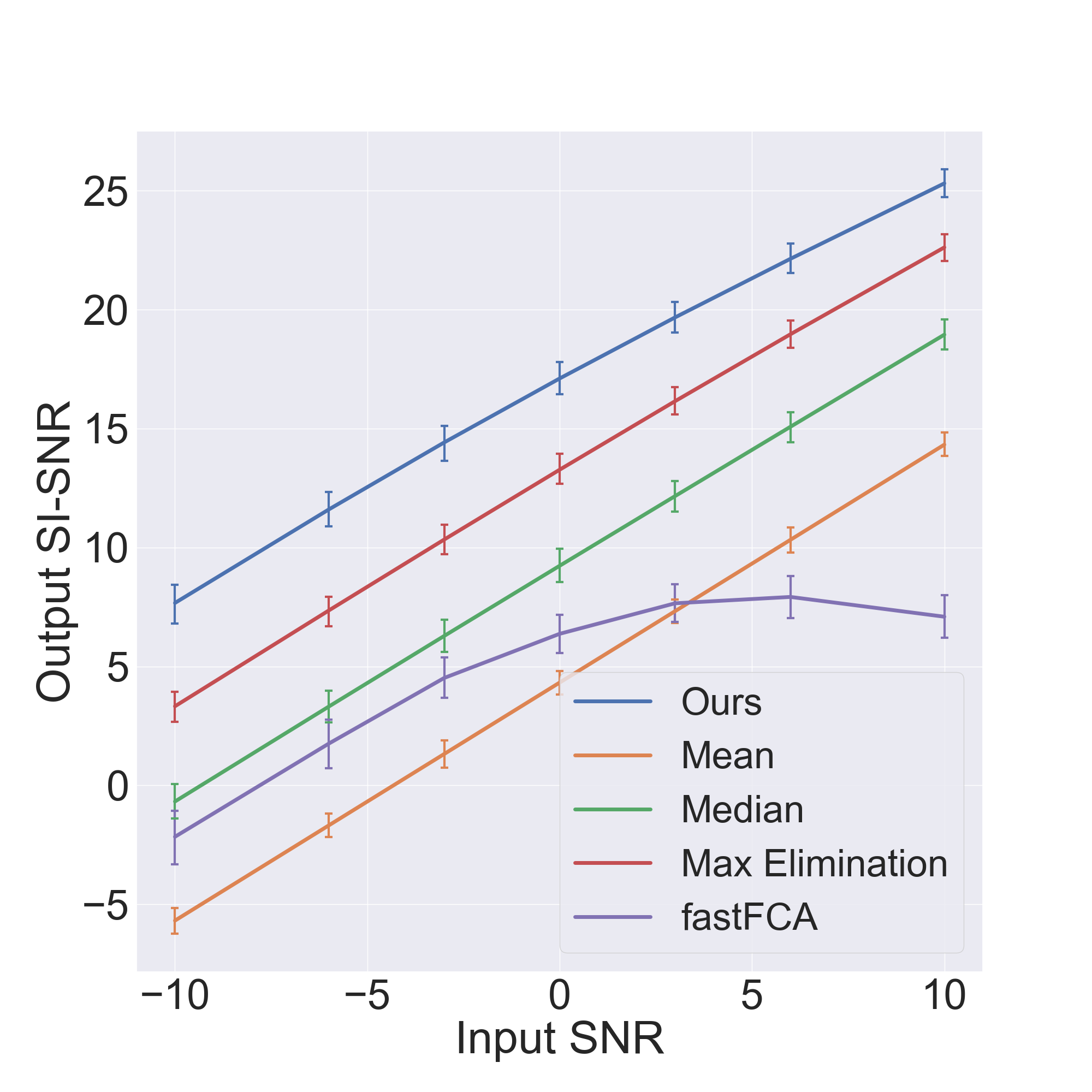}
         \caption{Speech vs. Noises (DNS) \label{fig:speech_and_different_DNS_noises}}
     \end{subfigure}
   
    \caption{Combining 5 synthetic noisy audio signals: Average SI-SNR of enhanced signal as a function of the SNR of the input signals, and $95\%$ confidence interval on 100 experiments. Methods compared: (i) Mean: Using the mean of all signals. (ii) Median: Replacing the mean magnitude with the median magnitude in each TF cell. (iii) Max Elimination \cite{MaxElimination:2017}: Removing the maximal magnitude in each TF cell. (iv) fastFCA: model the contribution of each source as a complex Gaussian distribution with zero mean. (v) Our Crowdsourced Enhancement, consistently having the best results.
    \label{fig:synthetic}}
\end{figure*}

\section{Experiments}
\label{sec:experiments}

We evaluate the proposed approach considering two different setups: (i) a synthetically generated dataset; and (ii) a dataset of real-world, 
user recordings collected from the web. The use of synthetic dataset allows us to evaluate the proposed approach in a controlled setting, exploring different noise levels and different types of noises. We also demonstrate that the proposed approach can generalize to user recording obtained from YouTube.

\begin{table*}[t!]
\caption{signal: speech vs. Noise: speech / signal: Speech vs. Noises (DEMAND) / Music vs. Noise: Speech . Same as Fig.~\ref{fig:synthetic}, but with PESQ and STOI as evaluation metric. The $95\%$ confidence interval ranges between $0.05-0.18$ and $0.01-0.02$ respectively }
 \resizebox{\textwidth}{!}{
  \centering  
  \begin{tabular}{c|l|l|l|l|l|l|l|l|l|l}
    \toprule
    \multirow{2}{*}{\textsc{SNR}} &
      \multicolumn{5}{c}{\textsc{PESQ}} &
      \multicolumn{5}{c}{\textsc{STOI}} \\
      \cline{2-6}
      \cline{7-11}
    & \textsc{Mean} & \textsc{Median} & \textsc{fastFCA} & \textsc{Max Elimi} & \textsc{Ours} & \textsc{Mean} & \textsc{Median} &  \textsc{fastFCA} & \textsc{Max Elimi} & \textsc{Ours} \\
    \toprule
    $-10 $ &  
    1.07 / 1.05 / 1.26 &
    1.16 / 1.06 / 1.34 &
    1.10 / 1.18 / 1.20 &
    1.25 / 1.12 / 1.59 &
    \textbf{1.61} / \textbf{1.22} / \textbf{2.18} &
    0.54 / 0.60 / 0.39 & 
    0.69 / 0.64 / 0.59 &
    0.60 / 0.73 / 0.45 & 
    0.77 / 0.73 / 0.70 & 
    \textbf{0.86} / \textbf{0.78} / \textbf{0.82}\\
    \hline
    $-6$ &
    1.10 / 1.09 / 1.21 &
    1.29 / 1.11 / 1.62 &
    1.23 / 1.33 / 1.31 &
    1.46 / 1.26 / 1.97 &
    \textbf{1.98} / \textbf{1.44} / \textbf{2.69} &    
    0.64 / 0.69 / 0.52 &
    0.78 / 0.73 / 0.70 &
    0.71 / 0.80 / 0.57 &
    0.84 / 0.81 / 0.79 &
    \textbf{0.91} / \textbf{0.85} / \textbf{0.87}\\
    \hline
    $-3 $&  
    1.14 / 1.14 / 1.30 & 
    1.44 / 1.19 / 1.91 &
    1.39 / 1.42 / 1.41 & 
    1.69 / 1.44 / 2.32 & 
    \textbf{2.32} / \textbf{1.69} / \textbf{3.04} &    
    0.70 / 0.76 / 0.61 &
    0.83 / 0.79 / 0.77 &
    0.78 / 0.84 / 0.65 &
    0.88 / 0.85 / 0.84 &
    \textbf{0.95} / \textbf{0.89} / \textbf{0.90}\\
    \hline
    $0 $& 
    1.23 / 1.24 / 1.51 & 
    1.64 / 1.33 / 2.26 &
    1.58 / 1.56 / 1.59 & 
    1.98 / 1.68 / 2.72 & 
    \textbf{2.68} / \textbf{2.00} / 
    \textbf{3.34} &    
    0.77 / 0.81 / 0.69 &
    0.87 / 0.84 / 0.82 &
    0.84 / 0.86 / 0.71 &
    0.91 / 0.89 / 0.88 &
    \textbf{0.95} / \textbf{0.92} / \textbf{0.92}\\
    \hline
    $3 $& 
    1.35 / 1.41 / 1.80 & 
    1.91 / 1.53 / 2.64 &
    1.78 / 1.64 / 1.78 & 
    2.32 / 1.98 / 3.06 & 
    \textbf{3.02} / \textbf{2.35} /\textbf{3.50}&    
    0.82 / 0.86 / 0.76 &
    0.90 / 0.88 / 0.86 &
    0.87 / 0.86 / 0.75 &
    0.93 / 0.92 / 0.91 &
    \textbf{0.96} / \textbf{0.94} / \textbf{0.94}\\
    \hline
    $6$ & 
    1.54 / 2.32 / 2.15 & 
    2.22 / 1.79 / 2.98 &
    2.00 / 1.70 / 1.96 & 
    2.68 / 2.32 / 3.36 & 
    \textbf{3.33} / \textbf{2.71} / \textbf{3.79} &    
    0.86 / 0.89 / 0.83 &
    0.93 / 0.91 / 0.90 &
    0.90 / 0.85 / 0.78 &
    0.95 / 0.94 / 0.93 &
    \textbf{0.97} / \textbf{0.95} / \textbf{0.96}\\
    \hline
    $10$ & 
    1.90 / 2.04 / 2.66 & 
    2.68 / 2.24 / 3.41 &
    2.20 / 1.88 / 1.97 & 
    3.14 / 2.82 / 3.68 & 
    \textbf{3.69} / \textbf{3.21} / \textbf{4.01} &    
    0.91 / 0.93 / 0.88 &
    0.90 / 0.94 / 0.92 &
    0.92 / 0.87 / 0.80 &
    0.97 / 0.96 / 0.94 &
    \textbf{0.98} \ \textbf{0.97} / \textbf{0.96}\\
\bottomrule
  \end{tabular}}
\end{table*}

\subsection{Datasets}
\label{sec:synthetic}
\noindent {\bf Synthetic Recordings.}
We artificially generated noisy inputs by mixing source and noise signals. As common audio source we use either music from the MUSDb18 benchmark~\cite{musdb18} or speech from the LibriSpeech corpus~\cite{panayotov2015librispeech}. All audio samples were resampled to 16kHz. Each common audio source signal $S$ is duplicate to k channels, while for each channel we add an independent noise. Each noise signal is multiplied by a different constant, $a_i$, which reflects the desired Signal-to-Noise Ratio (SNR) of the input signals. Formally, let $N_1, ...N_k$ be the noises added to each channel, the $a_i$ coefficients are computed as follows,
\begin{equation}
    \begin{aligned}
        &a_i = \sqrt{P(S)/(10^{SNR_{db}/10}\cdot P(N_i))},\\
        &P(S) = \sum_{n=1}^{\frac{len(S)}{\tau}}(\max_{t \in (n\tau,(n+1)\tau)}S(t))^2,
    \end{aligned}
\end{equation}
Where $\tau$ is a time interval, which was set to be 1 second.
We consider different types of noises such as speech, environmental noises, hammering, keyboard typing, dogs barking, etc. Speech data were obtained from the LibriSpeech corpus, while other types of noises were extracted from either DEMAND~\cite{thiemann2013diverse} or AudioSet~\cite{gemmeke2017audio}.

\noindent {\bf Real-world Recordings.}
We have collected real world user recordings of live music shows from YouTube. Multiple different clips were collected for each covered performance. As the clips were taken by independent users, we align and normalize all these recordings before processing. Overall, we collected $\sim$300 video recordings from $4$ different music shows.

\begin{figure*}[t!]     
     \begin{subfigure}[b]{0.32\textwidth}
         \centering
         \includegraphics[width=0.78\columnwidth, height=0.7\columnwidth]{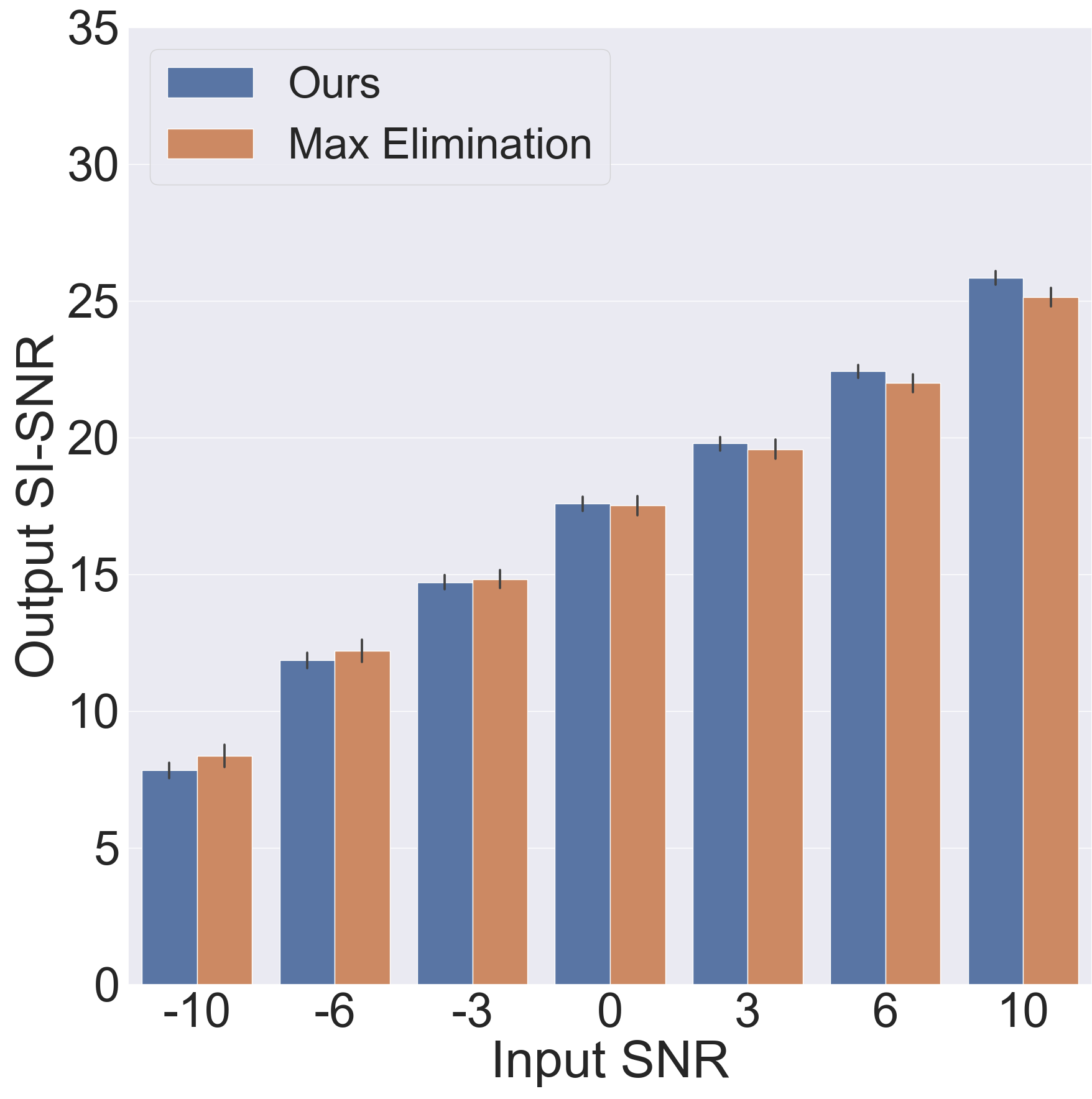}
         \caption{Three noisy inputs}
     \end{subfigure}~~~~     
     \begin{subfigure}[b]{0.32\textwidth}         
        \centering
         \includegraphics[width=0.78\columnwidth, height=0.7\columnwidth]{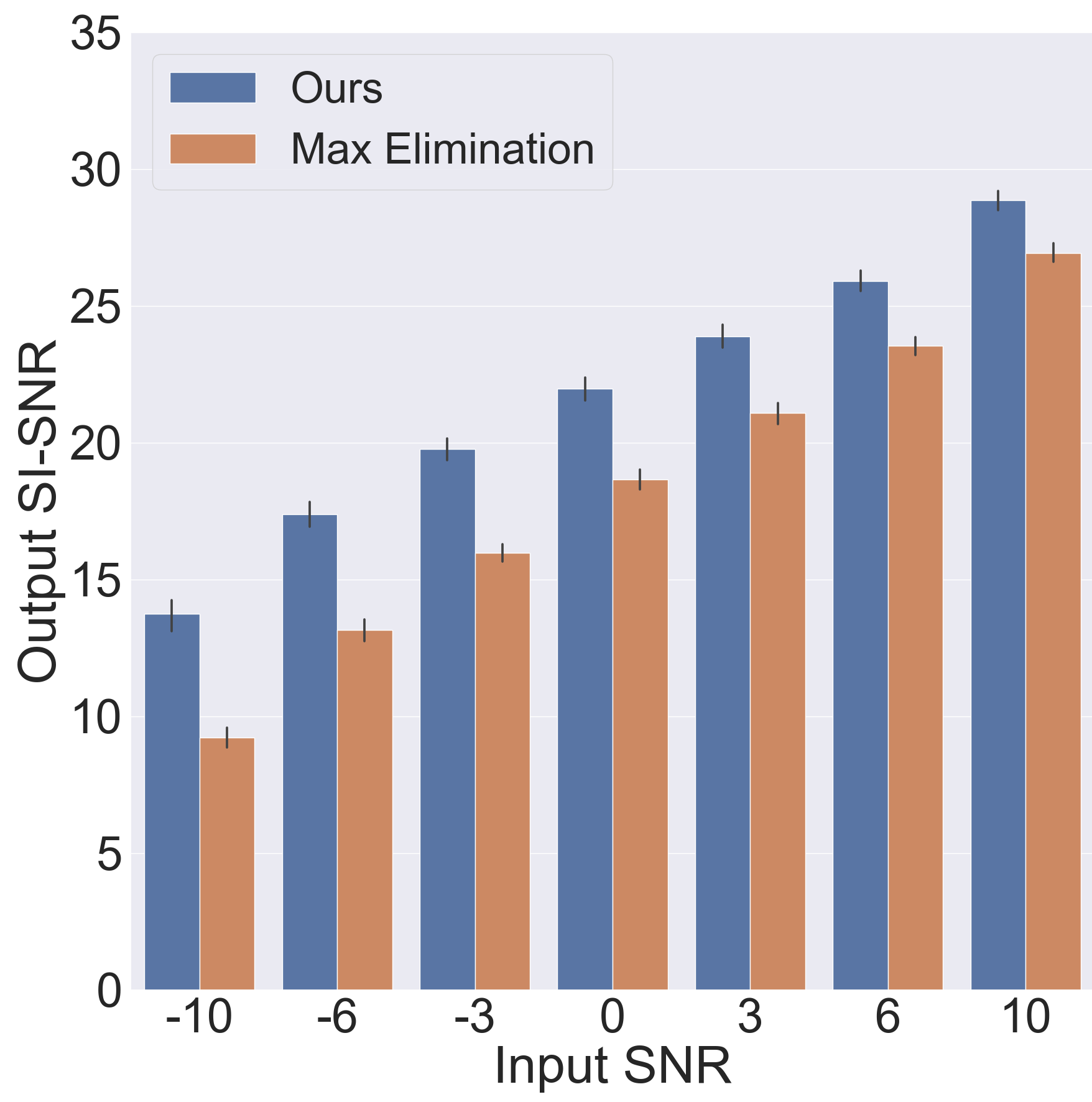}
         \caption{Five noisy inputs}
     \end{subfigure}~~~~ 
     \begin{subfigure}[b]{0.32\textwidth}
     \centering
         \includegraphics[width=0.78\columnwidth, height=0.7\columnwidth]{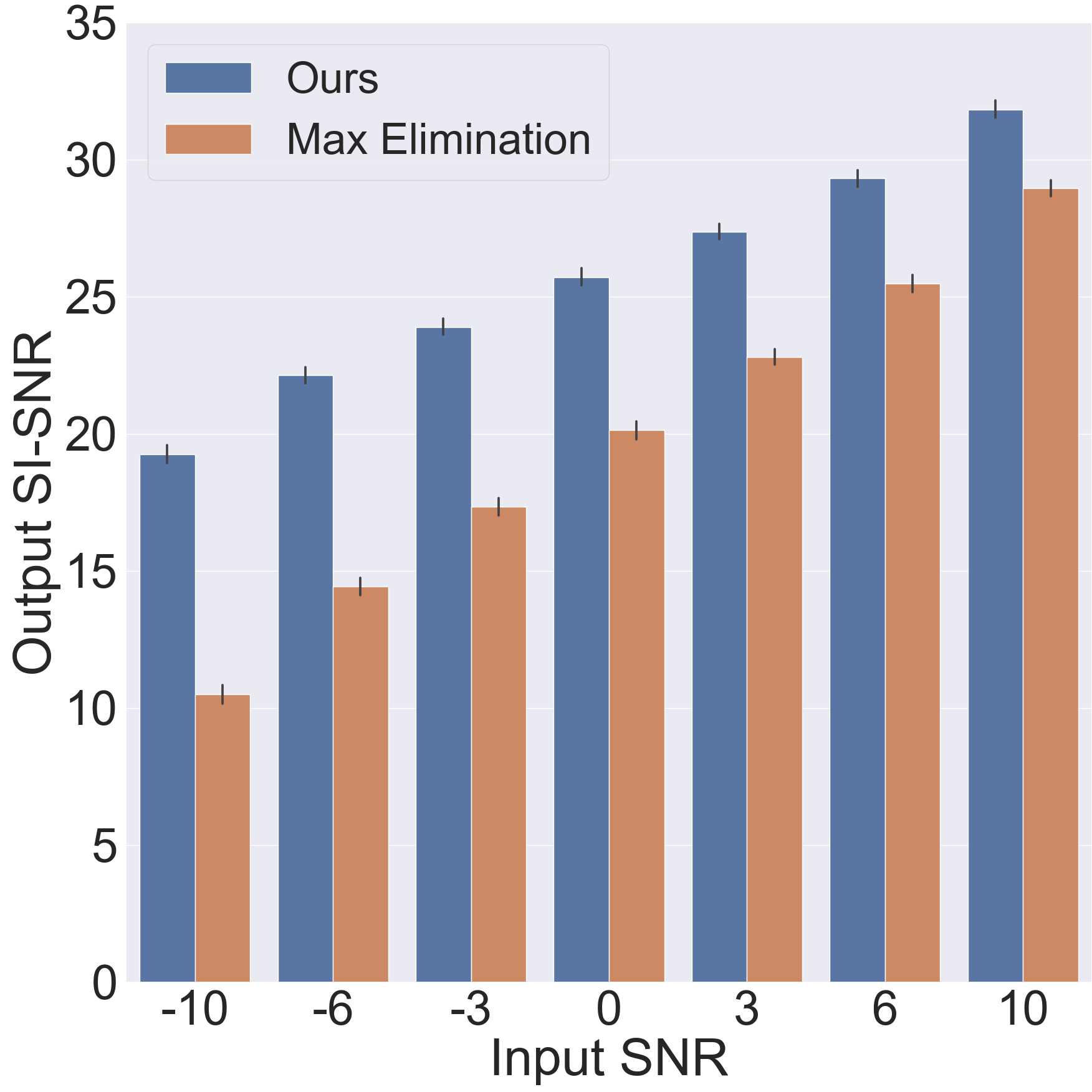}
         \caption{Ten noisy inputs}
     \end{subfigure}
    \caption{Average SI-SNR of enhanced signal, combining 3, 5, and 10 synthetic noisy audio signals. Source signal is music, and noise is speech. Max elimination \cite{MaxElimination:2017}, the best baseline, is compared with our Crowdsourced Enhancement. As expected, the benefit of our method over the baseline increases as more noisy signals are combined together.
    \label{fig:source_music+noise_speech_multi_m}}
\end{figure*}

\begin{figure*}[t!]
       \begin{subfigure}[b]{0.32\textwidth}
         \centering
         \includegraphics[width=0.9\columnwidth, height=0.7\columnwidth]{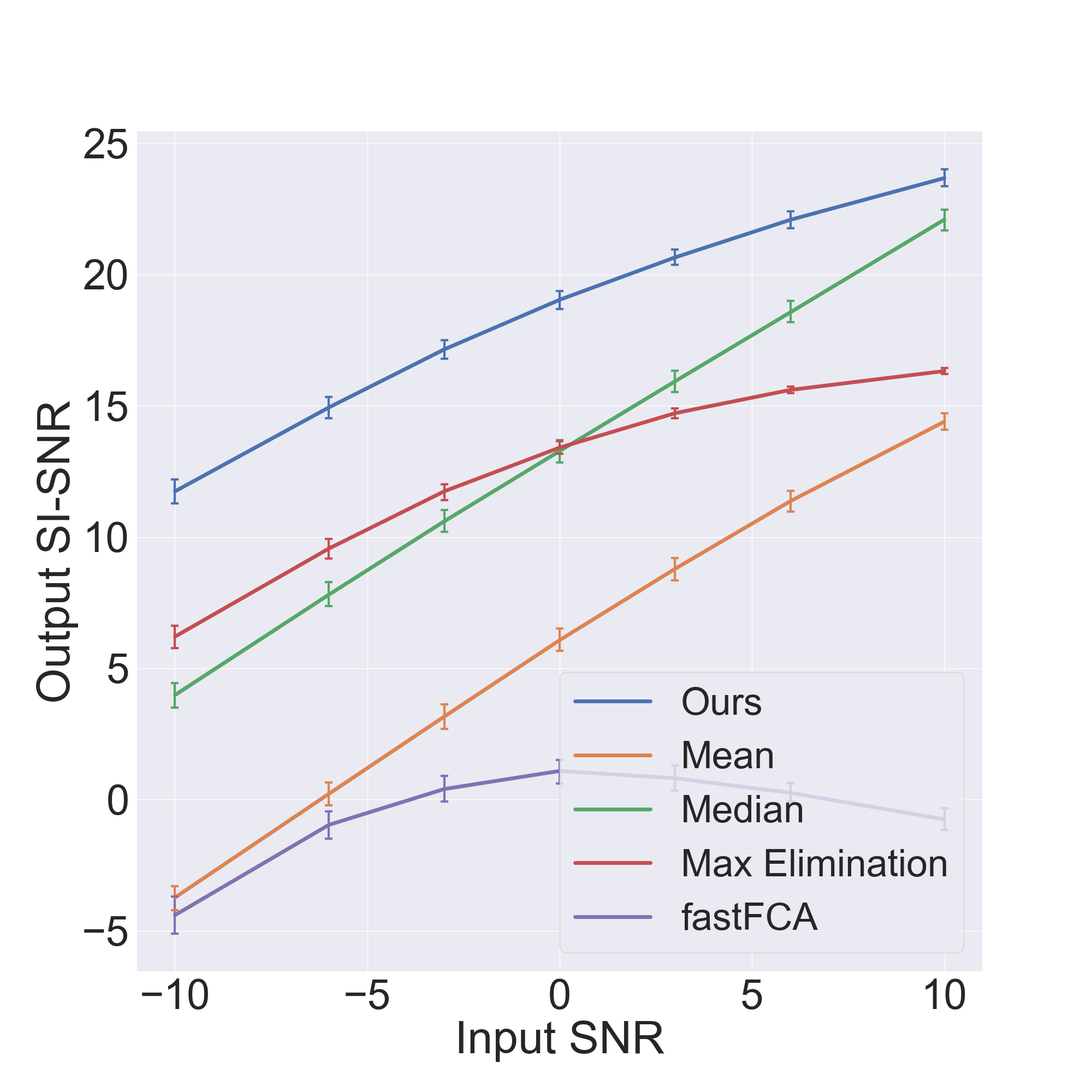}
         \caption{Signal: Speech vs. Noise: Speech \label{fig:speech_and_different_speech_noises1}}
     \end{subfigure} ~~~~
     \begin{subfigure}[b]{0.32\textwidth}
         \centering
         \includegraphics[width=0.9\columnwidth, height=0.7\columnwidth]{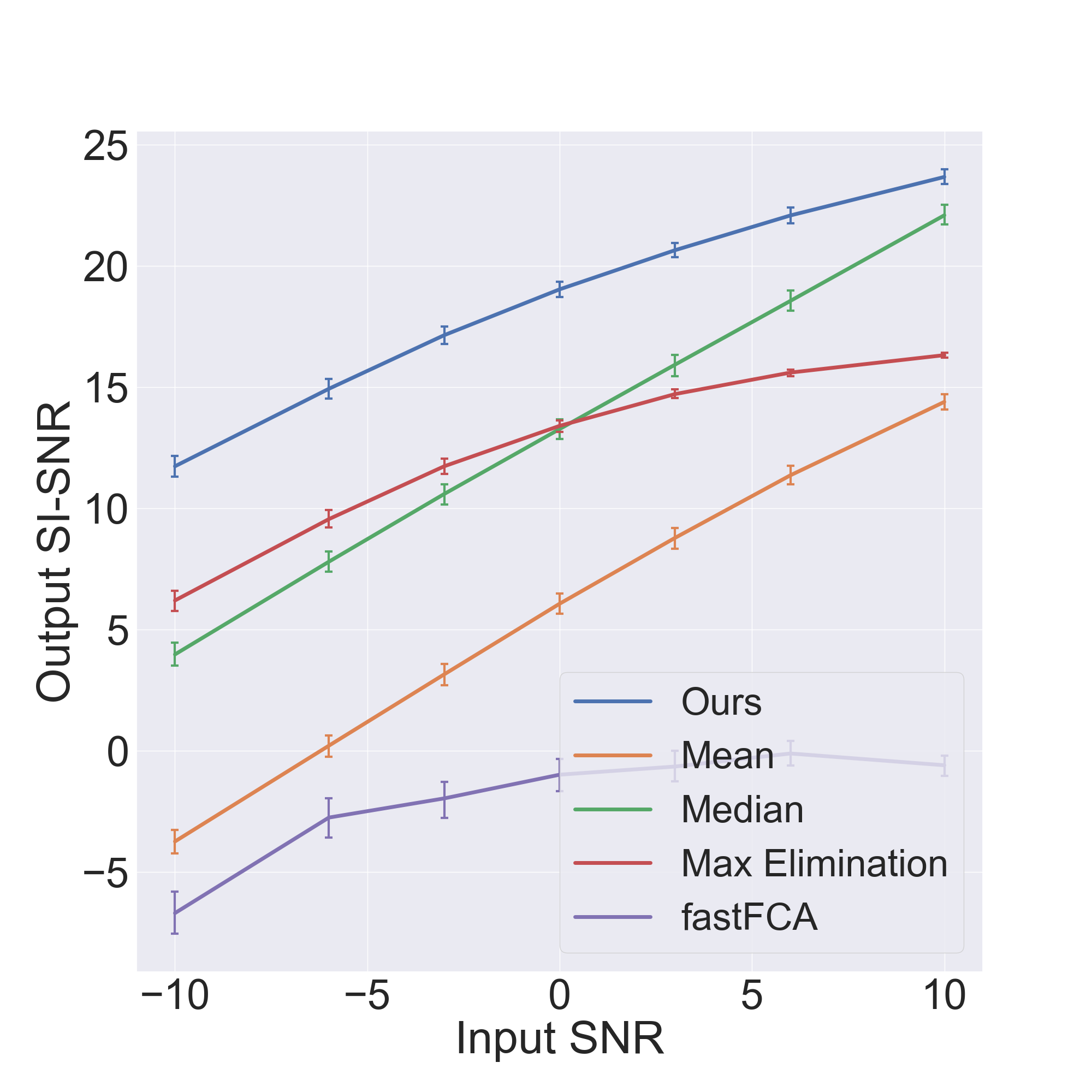}
         \caption{Speech vs. Noises (DEMAND) \label{fig:speech and different environmental noises1}}
     \end{subfigure}~~~~
     \begin{subfigure}[b]{0.32\textwidth}
         \centering
         \includegraphics[width=0.9\columnwidth, height=0.7\columnwidth]{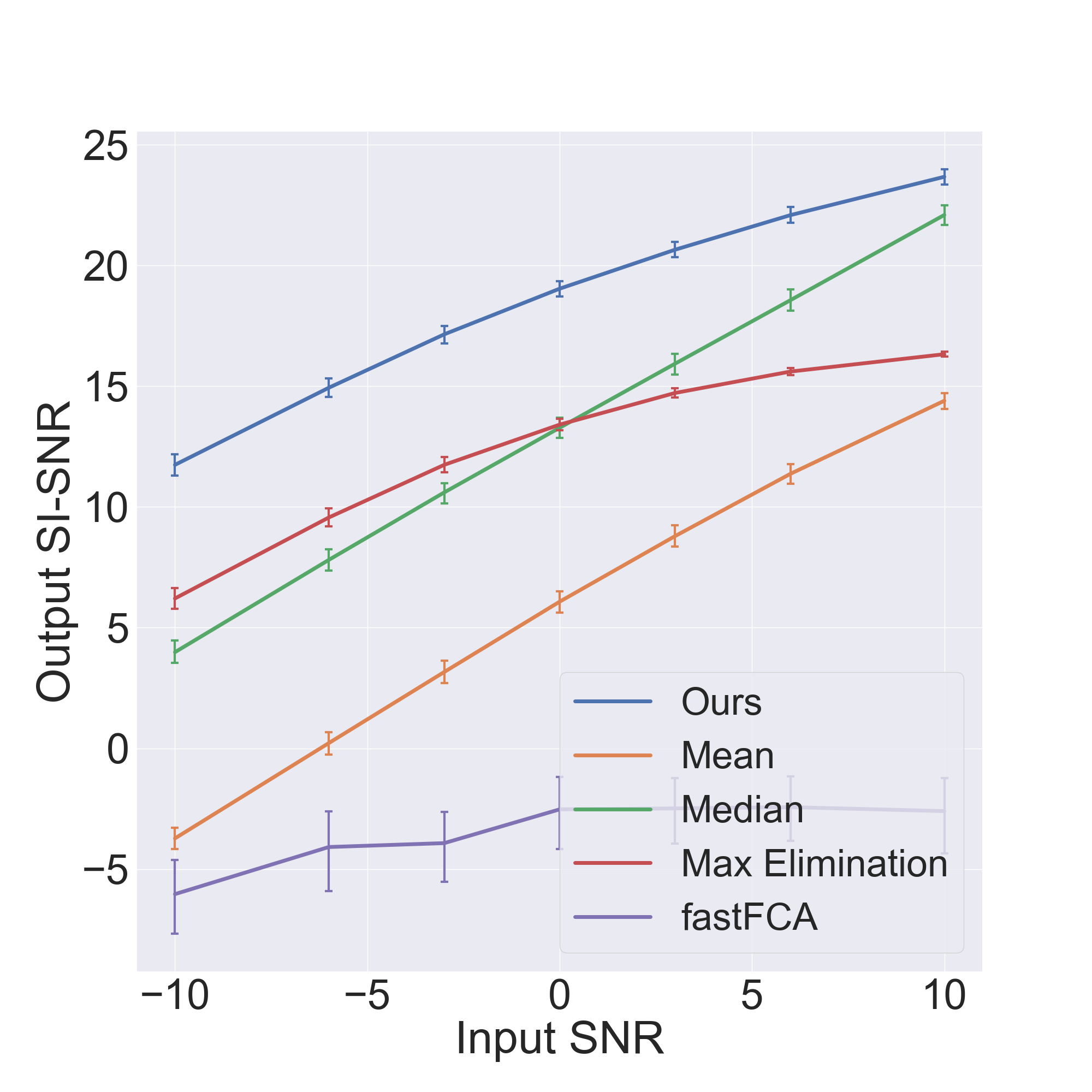}
         \caption{Music v.s Speech \label{fig:music_and_speech_noise1}}
     \end{subfigure}
   
    \caption{Same as Fig.~\ref{fig:synthetic}, but with simulated packet loss, where each noisy input signals also has a randomly placed one second of silence. Max Elimination, the best baseline under additive noise, fails in this case.
    \label{fig:synthetic2}}
\end{figure*}

\subsection{Baselines}

We evaluate the proposed method against four baselines. The first one, denoted as \textsc{Mean}, is constructed by taking the average of all input audio signals. The second baseline, denoted as \textsc{Median}, is constructed by computing the STFT of all signals and of the average signal, and replacing the magnitude of the average signal in each TF cell with the median magnitude of all input signals in that TF cell~\cite{Kim2013collaborative}. Another baseline is the \textsc{fastFCA}~\cite{ito2021joint}. In the time-frequency domain, each source contribution is modeled as a zero-mean Gaussian random variable whose covariance represents the source's spatial properties. We used the implementation described in \cite{ito2021joint}. The last baseline is the Maximum Component Elimination~\cite{MaxElimination:2017}, in which the magnitude of the average signal at each TF cell replaced by the average magnitude of all input signals in that TF cell after removing the maximal magnitude.

\subsection{Model Evaluation} 

To assess the quality of the reconstructed audio in relation to the reference signal the Invariant Signal-to-Noise Ratio (SI-SNR)~\cite{luo2018tasnet}, PESQ~\cite{rix2001perceptual}, \cite{miao_wang_2022_6549559}, and STOI~\cite{taal2011algorithm} were used as an objective methods, while we use the MUlti Stimulus test with Hidden Reference and Anchor (MUSHRA)~\cite{itu2001method} test as a subjective one. We conducted a human listening test using a web platform~\cite{schoeffler2018webmushra}, asking participants to rate the quality of recordings on a scale of 0 to 100 \cite{series2014method}.

\subsection{Results}
Results for the synthetic data can be seen on Figure~\ref{fig:synthetic} and Table 1 considering either music or speech as the source signal with various types of noises and SNR values. In all experiments we use $k=5$ sources. Notice, as this is a synthetic dataset, we have the perfect alignment, hence we skip the alignment process in this setting. We report the SI-SNR, STOI and PESQ metric between each of the methods against the clean target signal. Under each of the evaluated setups we extracted the enhanced signal using five synthesized noisy signals. Results suggest that the proposed method is significantly better than the evaluated baselines. This is more noticeable at low SNR values (e.g., -5, -10). Interestingly, when considering environmental noises from DEMAND, the gap between the proposed method and the evaluated baselines is smaller. 
In Figure~\ref{fig:source_music+noise_speech_multi_m} we compare our method to Max Elimination \cite{MaxElimination:2017}, considering different number of sources. Notice, the proposed method is superior to the Max Elimination method with an exception of three sources considering low SNR values. This implies that the proposed method can benefit from a large number of input sources.

Next, we experiment with a packet loss setting, where we assume random parts of each input signals may be missing. We inject a low energy white Gaussian noise in the missing periods, to prevent numerical issues with fastFCA. To simulate that, we randomly erase one second from each input signal independently. Results are presented in Figure~\ref{fig:synthetic2}. Results suggest the proposed method is superior to the evaluated baselines under this setting as well. Interestingly, as we go to higher SNR values, the Max Elimination method converges towards the mean. This can be explained as the Max Elimination considers one element less than the mean method and for high SNR values it is often not a noisy element. Notice that the median method is not affected by the packet loss as it will ignore it anyway.  

We perform subjective tests following the MUSHRA protocol~\cite{series2014method}, asking participants to rate the quality of recordings on a scale of 0 to 100. Obtained ratings: Max elimination~\cite{MaxElimination:2017}, the closest prior art, got $48.4 \pm 2.9$; our method got  $67.4\pm 2.6$, (mean $\pm$ $95\%$ confidence interval). This suggests that the proposed method is superior to the evaluated baselines also considering subjective metrics. Code, datasets, models and audio examples are available at the following link: \\ \url{https://shiranaziz.github.io/crowdsourced_audio_enhancement/}

\subsection{Recording from a Live Performance}
\label{sec:performance}
Finally, we evaluate the proposed approach on real recordings of live music shows collected from YouTube. As no ground truth is given when we enhance the crowdsourced recordings, the results can be examined on the website. 
\section{Conclusions}
We presented a simple and effective method for noise removal from crowdsourced recordings. The method examines individual time-frequency cells, and removes noisy input signals whose magnitude are outliers. The method can handle additive noise by removing outliers that are higher than the median signal, and can also handle silent moments (e.g., packet loss) by removing outliers lower than the median. We believe the development of simple and competitive baselines are crucial for constructing efficient solutions for real-world tasks. Although being simple, the proposed method improves over prior work, hence can be served as a new baseline for more complicated statistical methods which will be developed by the community in future work.

\bibliographystyle{IEEEtran}
\bibliography{refs}

\end{document}